\documentclass{PoS}
\usepackage{amsmath,amssymb}
\usepackage{float}
\title{Membrane Matrix models and non-perturbative checks of gauge/gravity duality} 

\ShortTitle{Membrane Matrix models}

\author{\speaker{Denjoe O'Connor}\\
    School of Theoretical Physics,\\ 
       Dublin Institute for Advanced Studies, \\
       10 Burlington Road, 
       Dublin 4, Ireland.\\
        E-mail: \email{denjoe@stp.dias.ie}}

\author{Veselin G. Filev,\\
    School of Theoretical Physics,\\ 
       Dublin Institute for Advanced Studies, \\
       10 Burlington Road, 
       Dublin 4, Ireland.\\
        E-mail: \email{vfilev@stp.dias.ie}
}

\abstract{We compare the bosonic and maximally supersymmetric membrane models.
  We find that in Hoppe regulated form the bosonic membrane
  is well approximated by massive Gaussian quantum matrix models. In contrast
  the similarly regulated supersymmetric membrane, which is equivalent
  to the BFSS model, has a gravity dual description. We sketch recent
  progress in checking gauge/gravity duality in this context.}

\FullConference{Proceedings of the Corfu Summer Institute 2015 "School and Workshops on Elementary Particle Physics and Gravity"\\
		1-27 September 2015\\
		Corfu, Greece}

\begin{document}
\section{Introduction}
\label{sec:introduction}
The proposal that there is a gravitational dual to gauged matter
systems represents a dramatic insight into certain non-perturbative
phenomena. The simplest known models with gravitational duals are
quantum mechanical models with extended supersymmetry and the simplest
of these is the model with maximal supersymmetry now known as the BFSS
model \cite{Banks:1996vh}. Though the model itself arose initially in
the context of supersymmetric quantum mechanical models
\cite{Baake:1984ie,Flume:1984mn,Claudson:1984th}, it subsequently emerged
from the matrix regularisation of membranes introduced by Hoppe
\cite{Hoppe:PhDThesis1982} and extended to the supermembrane in
\cite{de Wit:1988ig} and \cite{Townsend:1995kk}.\footnote{For a review see also \cite{Taylor:1999qk}.}  
It is this quantisation of membranes that is our focus here. The BFSS model 
was proposed as a non-perturbative formulation of M-theory which in 
the infinite matrix size limit is conjectured to capture the entire dynamics of M-theory.

In Hoppe's regularisation of membranes what remains of the original
diffeomorphism invariance of the membrane action is an $SU(N)$ gauge
symmetry.  The model has a single dimensionful coupling constant.
When placed in a thermal bath the coupling can be absorbed into the
fields and the temperature, to define a dimensionless temperature. At
high temperature the inverse of the dimensionless temperature plays
the r\^ole of a small parameter and the model is in a perturbative
regime. On the contrary at low temperature the the model becomes
strongly coupled.

In the low temperature regime the bosonic model turns out
\cite{Filev:2015hia} to be well described by a massive quantum matrix
model while the supersymmetric model is described by a dual geometry
\cite{Itzhaki:1998dd} which is a solution to IIA supergravity and can
be lifted to a solution to 11-d supergravity. For the thermal system
the gravity dual has a black hole whose Hawking-temperature is that of
the thermal bath.

A more complicated example that was designed to describe the M2-brane
in a background longitudinal M5-brane is the Berkooz-Douglas
model~\cite{Berkooz:1996is}.  The model also arises naturally in
string theory as the effective low energy description of a D0/D4-brane
system. The D0-branes give rise to the adjoint fields of the pure BFSS
model, while the D4-branes are described by fundamental
hypermultiplets. In the large $N$ limit at strong 't Hooft coupling
the model has a supergravity dual description. The best understood
regime of this duality is when the number of D4-branes is much smaller
than the number of D0-branes, which from a field theory point of view
corresponds to the quenched approximation. In this limit the D0-branes
are described by the supergravity dual of the BFSS model, while the
D4-branes are treated as Dirac--Born--Infeld probes. The model was successfully simulated in ref.~\cite{Filev:2015cmz}, where excellent agreement between field theory and supergravity has been found.  

We will review some recent progress in checking this duality and
explain its significance to a non-perturbative formulation of
M-theory.\footnote{For a recent review on the subject we refer the reader to ref.~\cite{Hanada:2016jok}.}

\section{Hoppe Regularised Membranes}
The Membrane action, with metric signature $(-,+,..+)$, in Polyakov form
is given by 
\begin{equation}
S=-\frac{T}{2\Lambda}\int d^3\sigma\sqrt{-h}\left(h^{\alpha\beta}\partial_\alpha X^\mu\partial_\beta X^\nu\eta_{\mu\nu}-\Lambda\right)\qquad\hbox{with }\quad T=\frac{1}{(2\pi)^2{\it l}_p^3}\; .
\label{SMembrane}
\end{equation}
Varying with respect to $h_{\alpha\beta}$ one obtains
\begin{equation}
  \frac{1}{2}h_{\gamma\delta}(h^{\alpha\beta}G_{\alpha\beta}-\Lambda)=G_{\gamma\delta}\;,
  \quad\hbox{tracing gives }\quad h^{\alpha\beta}G_{\alpha\beta}=3\Lambda\; ,\quad \hbox{so } \Lambda\; h_{\gamma\delta}=G_{\gamma\delta}\; . 
\end{equation}
Then substituting into (\ref{SMembrane}) gives the standard Nambu-Goto form
\begin{equation}
S=-T\int d^3\sigma \sqrt{-G}\; .
\end{equation}

Alternatively, instead of going to the Nambu-Goto form of the action, we can
gauge away some of its components of $h_{\alpha\beta}$ and
attempt to solve the resulting constraints on the embedding coordinates.
When the membrane topology is restricted to 
${\mathbb R}\times\Sigma$ we can use the gauge
$h_{0i}=0$ and $h_{00}=-\frac{4}{\Lambda\rho^2}{\rm det}(G_{ij})$
so we have the constraint
\begin{equation}
  \partial_t X^\mu\partial_i X^\nu\eta_{\mu\nu}=0\quad
\hbox{and}\quad \Lambda\; h_{00}=  \partial_tX^\mu\partial_tX^\mu\eta_{\mu\nu}=-\frac{4}{\rho^2}{\rm det}(G_{ij})\ .
\end{equation}
Using light-cone coordinates with $X^{\pm}=(X^0\pm X^{D-1})/\sqrt{2}$ 
and choosing $X^+=\tau$ we see that the constraints take the form
\begin{equation}
  \partial_i X^{-}=(\partial_\tau X^a)(\partial_i X_a)
  \quad \hbox{and}
  \quad \partial_\tau X^-=\frac{1}{2}\partial_\tau X^a\partial_\tau X_a
  + \frac{2}{\rho^2} {\rm det}(G)
\end{equation}
with the action
\begin{equation}
S=\frac{T\rho}{4}\int dt d^2\sigma\left(\dot X^\mu \dot X^\nu\eta_{\mu\nu}
-\frac{4}{\rho^2}{\rm det}(G_{ij})\right)\ .
\nonumber\end{equation}

An important observation is that in two dimensions ${\rm det}(G_{ij})$ can be rewritten using 
$\{f,g\}=\epsilon^{ij}\partial_i f\partial_j g$ so that 
\begin{equation}
S=\frac{T\rho}{4}\int dt d^2\sigma \left(\dot X^\mu \dot X^\nu\eta_{\mu\nu}
-\frac{4}{\rho^2}\{X^\mu,\,X^\nu\}^2\right)
\nonumber\end{equation}
and the constraints become
\begin{eqnarray}
\dot{X}^\mu\partial_i X_\mu=0\implies \{\dot{X}^\mu,\,X_\mu\}=0
\quad{\rm and}\quad\dot{X}^\mu\dot{X}_\mu =-\frac{2}{\rho^2}\{X^\mu,\,X^\nu\}\{X_\mu,\,X_\nu\}\, .
\end{eqnarray}
In light-cone coordinates the Lagrangian is linear in the momentum $P_-$
and a Legendre transform to the Hamiltonian gives
\begin{equation}
S=-T\int\sqrt{-G}\longrightarrow H=\int\left(\frac{1}{\rho T}P^aP^a+\frac{T}{2\rho}\{X^a,\,X^b\}^2\right)\; ,
\nonumber
\end{equation}
with the remaining constraint $\{P^a,\,X^a\}=0$.

This model is still not fully tractable, however Hoppe then made the observation that if he treated the membrane surface as a quantum phase space
one could use a matrix regularisation of the membrane. 
In this scheme functions on the membrane world-volume at fixed time, 
$f(\sigma^1,\, \sigma^2)$  are replaced by $N\times N$ matrices,  
$f\rightarrow F$, with the matrices providing a discrete approximation to 
the corresponding functions.
This is the same
procedure as is used in the fuzzy approach \cite{Balachandran:2005ew} with
the significant difference that the geometry of $\Sigma$, the membrane surface,
is lost.

The quantum Hamiltonian then reads
\begin{equation}
{\rm H}=-\frac{1}{2}\nabla^2-\frac{1}{4}\sum_{i,j=1}^p{\rm Tr}[X^i,\,X^j]^2\nonumber
\end{equation}
and the constraint requires that observables are restricted to
$SU(N)$ invariants.
The model describes a quantised ``fuzzy'' relativistic membrane in $p+1$ dimensions.

The Euclidean finite temperature action for the model is 
\begin{equation}\label{BosS}
S_b = \frac{1}{g^2}\,\int_0^{\beta}dt\,{\rm tr}\left\{\frac{1}{2}({\cal D}_t {X^i})^2-\frac{1}{4}[X^i,\,X^j]^2\right\}\ .
\end{equation}
where ${\cal D}_t {X^i}=\partial_t X^i+[A,\,X^i]$ and $\beta$, the period of 
the $S^1$, is the inverse temperature. Since the theory is one dimensional
the only physical content of the gauge field is its holonomy around the $S^1$.
It is also the high temperature limit of a 
$1+1$ dimensional ${\cal N}=8$ supersymmetric Yang-Mills theory
on ${\bf R}\times S^1$  where $\beta$ is now the period of 
a spatial $S^1$ ({\bf not the inverse temperature}) and
the fermions drop out due to their anti-periodic 
boundary conditions at finite temperature~\cite{Aharony:2004ig}.

This model has been studied in some detail both non-perturbatively~\cite{Kawahara:2007fn,Filev:2015hia} and using a $1/p$
expansion~\cite{Filev:2015hia, Mandal:2011hb,  Azuma:2014cfa}
where $p$ is the spatial dimension into which the membrane is embedded.
It was found that as the temperature is decreased the model first undergoes a
2nd order deconfining-confining phase transition into a phase with
non-uniform but gapless distribution for the holonomy.  As the
temperature is further decreased there is a 3rd order transition to a
gapped holonomy with a quadratic decrease in the internal energy to a
constant value for lower temperatures. The high temperature expansion
of the model was developed in~\cite{Kawahara:2007ib}.
In \cite{Filev:2015hia}  it was found that the low temperature phase of
the model has an effective description in terms of free massive
scalars which captures many of the finite temperature features of the model
including one of its two phase transitions.

For the purposes of the discussion here the zero temperature aspects of the model are most interesting. In this case, the gauge field which in the static gauge
enters only as a holonomy at finite temperature, can be completely gauged away
and the model
simplifies. Furthermore, at zero temperature the correlator:
\begin{equation}
\left\langle{\rm Tr}\left(X^1(0)\,X^1(t)\right)\right\rangle\propto e^{-m\,t} +\dots\ ,
\end{equation}
captures the gap $m = E_1-E_0$ of the theory. To calculate the gap in the discrete theory, we periodically identify the time direction with period $\beta$:
\begin{equation}\label{corel}
\left\langle{\rm Tr}\left(X^1(0)\,X^1(t)\right)\right\rangle =A\,( e^{-m\,t} +e^{-m(\beta -t)}) \ ,
\end{equation}
Note that although formally $\beta$ is the same parameter that we have
at finite temperature, since we set the holonomy to zero here its
meaning is just a periodic coordinate as opposed to inverse temperature. Our
result for the correlator for $N=30$, $\beta=10$ and lattice spacing
$a=0.25$ is presented on the left in Figure \ref{fig:4}. The fitting
curve is given by equation (\ref{corel}) and when we perform a two parameter
fit we obtain $A\approx 7.50\pm 0.2$ and
$m\approx (1.90\pm.01)\,\lambda^{1/3}\,$. However, for Gaussian scalar
fields of mass $m$ we have $A=\frac{N}{2m(1-{\rm e}^{-\beta m})}$ and
performing a one parameter fit for $m$ yields $m=1.965\pm0.007$
and $A=7.63\pm 0.03$. On the
right we have presented a plot of the eigenvalue distribution of one
of the matrices for the same parameters. The fitting curve represents
a Wigner semicircle of radius $R_\lambda\approx 1.01$. The fact that the
theory is gapped and that the eigenvalue distribution is a semicircle
suggests that that at low temperate the model has an effective action:
\begin{figure}[t] 
   \centering
   \includegraphics[width=2.9in]{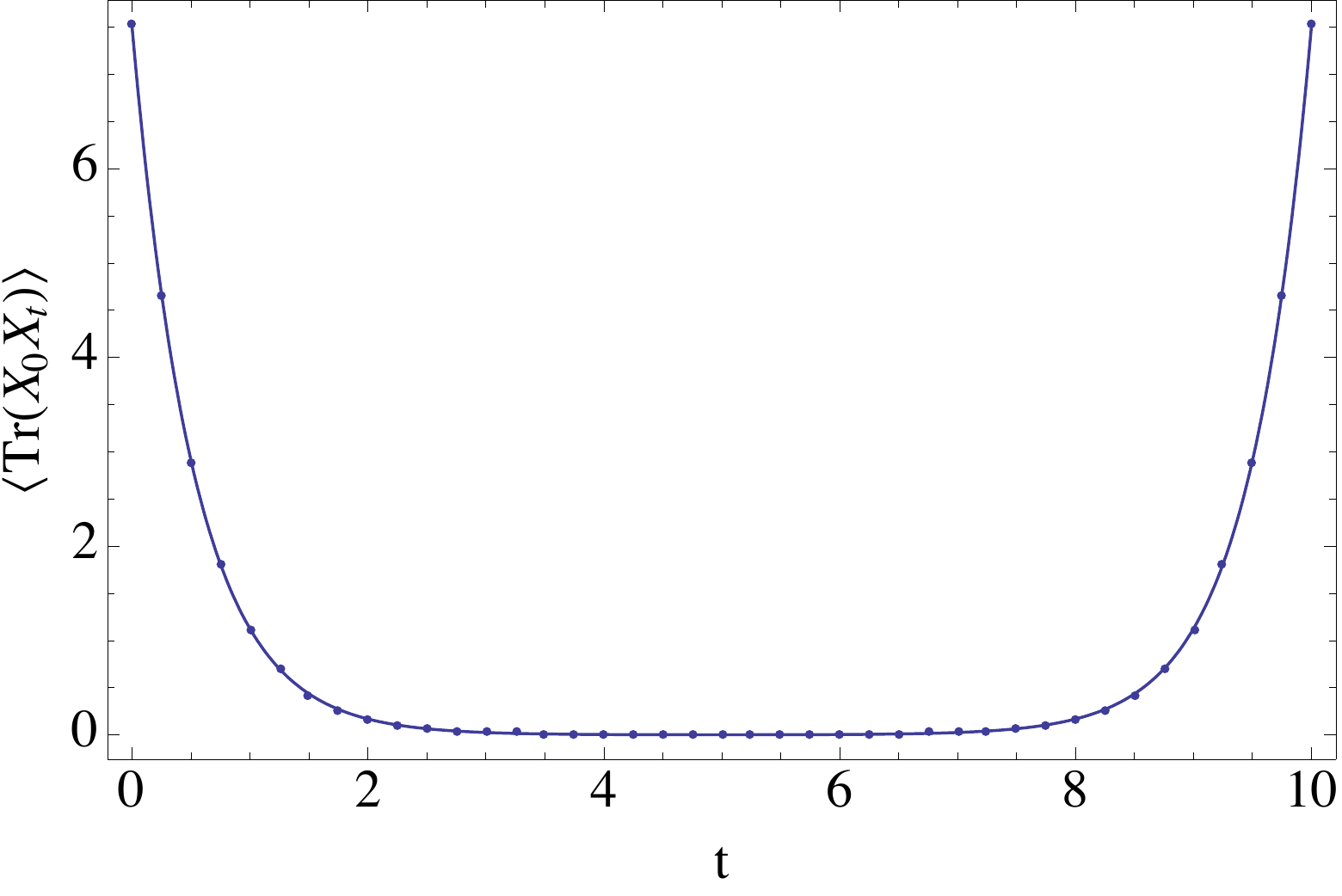} 
   \includegraphics[width=2.9in]{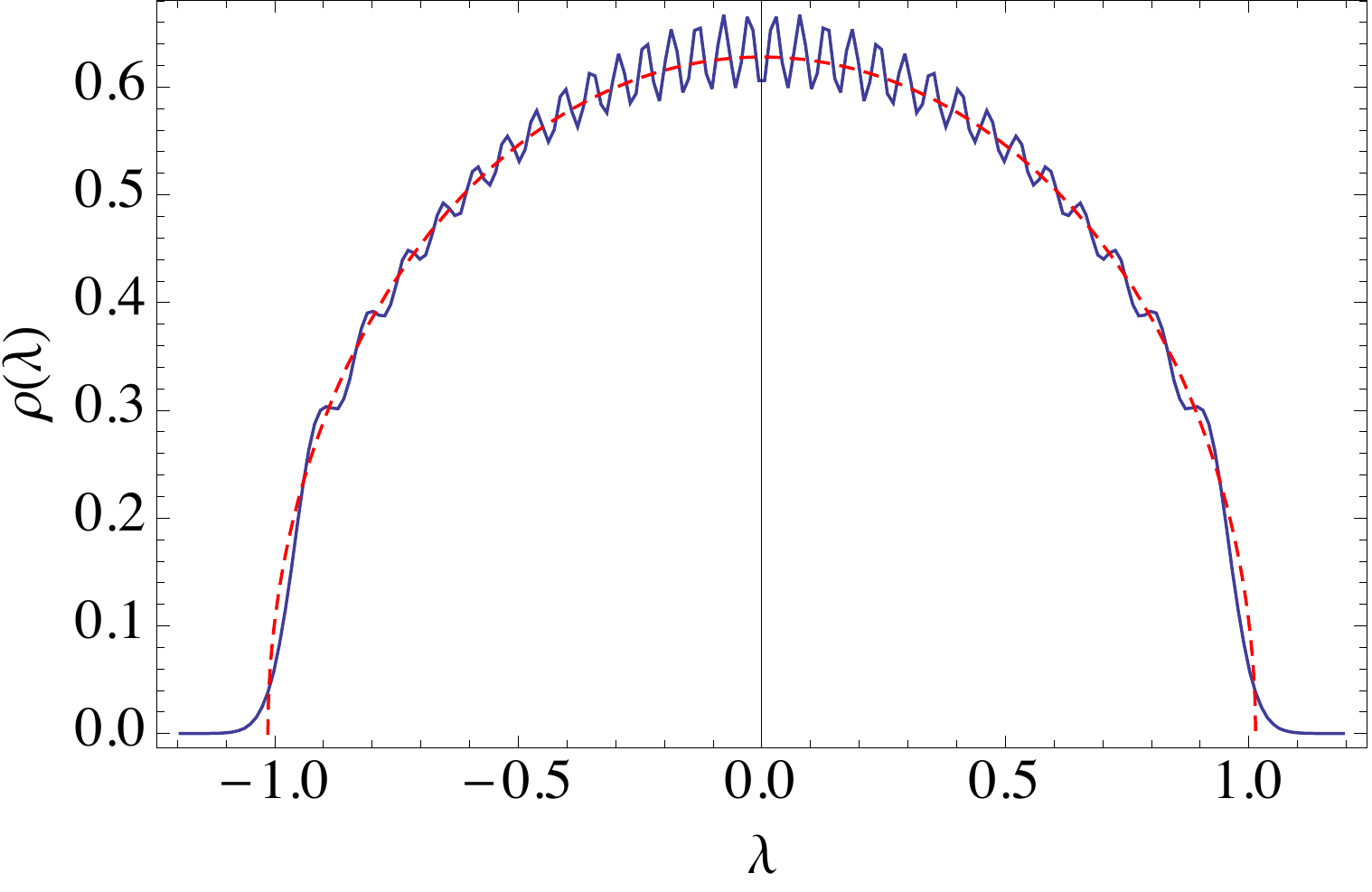} 
   \caption{\small {\it On the left:} A plot of the correlator
     $\left\langle{\rm Tr}\left(X^1(0)\,X^1(t)\right)\right\rangle$ for
     $N=30$, $\beta =10$ and lattice spacing $a=0.25$. The fitting curve is
     given by equation (2.9) 
     with $A=\frac{N}{2m(1-{\rm e}^{-\beta m})}$ and with parameters
     $m \approx (1.965\pm.007)\lambda^{1/3}\,$.
     {\it On the right:} A plot of the eigenvalue distribution of
     one of the scalars for the same parameters. The fitting curve
     represent a Wigner-semicircle of radius $R_\lambda\approx 1.01\,$.}
   \label{fig:4}
\end{figure}
\begin{equation}
S_{\rm eff}=N\int\limits_{-\infty}^\infty dt\,{\rm Tr}\left( \frac{1}{2}\left(\dot X^i\right)^2+\frac{1}{2}m^2 \left(X^i\right)^2\right)
\label{GaussianModel}
\end{equation}
for each of the matrices $X^i$. It is well known \cite{Brezin:1977sv}
that for the action (\ref{GaussianModel}) the eigenvalue distribution
of $X$ is given by a Wigner semicircle of radius:
\begin{equation}
R_\lambda =\sqrt{\frac{2}{m}}\approx 1.009\pm.002,
\end{equation}
where we have substituted $m \approx 1.965\pm .007$.
This agrees nicely (within errors) with the result for $R_\lambda\approx 1.01$
obtained by fitting the actual distribution. It is also in excellent
agreement with the large $p$ theoretical prediction of \cite{Mandal:2011hb},
\begin{equation}
R_\lambda(p)=\sqrt{\frac{2}{p^{1/3}}}\left(1+\frac{1}{p}\left(\frac{7\sqrt{5}}{30}-\frac{9}{32}\right)+\cdots\right)\simeq 1.0068
\end{equation}

Numerical simulations of the gauged massive Gaussian model show
that it too has a phase transition (there is only one rather than
the two of the full model) and that when the mass is tuned
to the value of the full model the transition temperature is
almost identical~\cite{Filev:2015hia}.

That the Hoppe quantised bosonic membrane has a mass gap was clear from
the analysis of \cite{deWit:1988ct}, the new ingredient here is
putting a value on this mass and demonstrating how closely
the model is to a gauged massive Gaussian model.

\subsection{Quantum Gravity from matrices}

It has been argued \cite{Doplicher:1994tu} that at short distances due
to quantum gravity the  spatial coordinates, $X^a$ should not
commute i.e. $[X^a,\,X^b]\neq 0$ somewhat in analogy with the non-commutativity of
phase space, but putting this non-commutativity to a spacetime independent
quantity, e.g. $[X^a,\,X^b]=i\,\theta^{ab}$ breaks rotational invariance.
One would wish to have the non-commutativity be a feature of the high energy
theory and disappear at low energies.  If one takes each $X^a$
to be an $N\times N$ matrix (as in matrix mechanics) and tries
\begin{equation}
  H_0={\rm Tr}\left(\frac{1}{2}\sum_{a=1}^{p}\dot{X}^a\dot{X}^a-\frac{1}{4}\sum_{a,b=1}^{p}[X^a,\,X^b][X^a,\,X^b]\right)
\end{equation}
one can see that the bottom of the potential has $[X^a,\,X^b]=0$ and one
might hope that at low energy the matrices would effectively
commute\footnote{It is interesting to see what would be the physics of the
  ensemble of eigenvalues if quantum effects did not lift the minimum
  drastically. This was studied in \cite{Filev:2014qxa} where it was
  found that for an ensemble of $p$ commuting matrices, $X^a$, in a
  quadratic potential, the $X^a$ are concentrated on a sphere for
  $p\ge4$.}. However, as we saw above the model becomes
massive at low energies and the minimum
is pushed upwards so that the non-commutativity does not disappear.
The $X_a$ are always described (see Figure \ref{fig:4}) by a Wigner
semi-circle.

Following arguments similar to those above, Polchinski
argues \cite{Polchinski:2014mva} that one might suspect that the model
still has something to do with quantum gravity and that the missing
ingredient is supersymmetry.  The model can be made supersymmetric
by adding a fermionic degree of freedom for each bosonic degree of
freedom so that
\begin{equation}
  H_1={\rm Tr}\left(\frac{1}{2}\sum_{a=1}^{p}\dot{X}^a\dot{X}^a-\frac{1}{4}\sum_{a,a=1}^{p}[X^a,\,X^a][X^a,\,X^a]+\frac{1}{2}\Theta^T\gamma^i[X^a,\,\Theta]\right)
\end{equation}
with the Fermions quantised as
$\{{\Theta_\alpha}_{i}^{j},\,{\Theta_\beta}_{k}^{l}\} =\delta_{\alpha\beta}\delta_{ik}\delta^{jl}$, i.e. as Clifford algebra elements or equivalently as
fermionic Majorana oscillators.
It was observed in \cite{Baake:1984ie} that one can only match fermions with
bosons if $p=2,3,5$ or $9$.

The potential is invariant under the global $SU(N)$ transformations
$X^a\rightarrow UX^a U^{-1}$, so promoting this symmetry to a gauge symmetry
with $U=U(t)$ and $\dot{X}^a\rightarrow D_t X^a=\dot{X}^a+i[A,\,X^a]$
where $A$ is the $SU(N)$ gauge field gives the matrix regularisation of
the supermembrane \cite{de Wit:1988ig}.
The constraint of supersymmetry then means that supersymmetric
membranes only exist in spacetime dimensions $4$, $5$, $7$ and $11$.

\section{The BFSS model}

Matrix supermembranes propagating in $p + 2$ dimensions coincide with
$p + 1$-dim $SU(N)$  supersymmetric Yang-Mills theory dimensionally reduced
to one dimension (only time  dependence). The BFSS model, $p = 9$, 
also describes a system of N interacting D0 branes.

\subsection{The Hamiltonian Formulation}
  
The $16$ supercharges:
\begin{equation}
Q_\beta=Tr\left(\frac{1}{2}\Theta_\alpha\gamma^a_{\alpha\beta}P_a+\frac{i}{4}\Theta_\alpha\gamma^{ab}_{\alpha\beta}[X_a,X_b]\right)\nonumber
\end{equation}
\begin{equation}
\{Q_\alpha,Q_\beta\}=\delta_{\alpha\beta}{\cal H}+\gamma^a_{\alpha\beta}Tr(X^aJ)\nonumber
\end{equation}
give the Hamiltonian:
\begin{equation}
{\rm H}=Tr\left(\frac{1}{2}P^aP^a-\frac{1}{4}[X^a,X^b]^2+\frac{1}{2}\Theta^T\gamma^a[\Theta,X^a]\right)\nonumber\ ,
\end{equation}
where $J$ is the generator of $SU(N)$ and is zero on physical states
\begin{equation}
J=i[P^b,X^b]+\Theta_\alpha\Theta_\alpha-\delta_{\alpha\alpha}\frac{N^2-1}{2N}\; .
\nonumber
\end{equation}
The 16 fermionic matrices $\Theta_{\alpha}=\Theta_{\alpha A}t^A$ 
are quantised as ${\Theta_{\alpha A},\Theta_{\beta B}}=2\delta_{\alpha \beta}\delta_{AB}$. 
The $\Theta_{\alpha A}$ are $2^{8(N^2-1)}$ and the Fermionic Hilbert space
is $${\cal H}^{F}={\cal H}_{256}\otimes\cdots\otimes{\cal H}_{256}$$ with 
${\cal H}_{256}={\bf 44}\oplus{\bf 84}\oplus{\bf 128}$
suggestive of the graviton, anti-symmetric tensor and gravitino of 
$11$-dimensional supergravity.
For an attempt to find the ground state if this Hamiltonian see ref.~\cite{Hoppe:2008uc}.

\subsection{Lagrangian formulation}

The easiest way to obtain the
BFSS matrix model is via dimensional reduction of ten dimensional
supersymmetric Yang-Mills theory down to one dimension. The resulting
reduced ten dimensional action is given by

\begin{eqnarray}\label{BFSS 10Mink}
S_M && =\frac{1}{g^2}\int dt \,{\rm Tr}\left\{\frac{1}{2}({\cal D}_0X^i)^2 +\frac{1}{4}[X^i,X^j]^2
-\frac{i}{2}\Psi^T C_{10}\,\Gamma^0D_0\Psi +\frac{1}{2}\Psi^T C_{10}\,\Gamma^i[X^i,\Psi]\right\}\ ,
\end{eqnarray}
where $\Psi$ is a thirty two component Majorana--Weyl spinor,
$\Gamma^\mu$ are ten dimensional gamma matrices and $C_{10}$ is the
charge conjugation matrix satisfying $C_{10} \Gamma^{\mu}C_{10}^{-1} =
-{\Gamma^{\mu}}^T$.

\section{The gravity dual}

Since the BFSS model describes the dynamics of D0-branes gauge/gravity dualtiy gives
predictions for the strong coupling regime of the theory. To access
these predictions one needs to look at the supergravity. Also, since
the supermembrane is meant as the basic ingredient of M-theory and in
analogy with string theory where quantised superstrings gives
10-dimensional supergravities as their low energy theories, the
supermembrane is expected to give 11-dimensional supergravity as its
low energy theory. It turns out that the dual geometry for
the BFSS model can be lifted to a solution to 11-dimensional
supergravity. We briefly describe the solution in this context.

The bosonic action for eleven-dimensional supergravity is given by:
\begin{equation}
S_{11D}=\frac{1}{2\kappa_{11}^2}\int \left[\sqrt{-g}R-\frac{1}{2}F_4\wedge *F_4-\frac{1}{6} A_3 \wedge F_4\wedge F_4\right]\ ,
\end{equation}
where $2\kappa_{11}^2=16\pi G_N^{11}=\frac{(2\pi l_p)^9}{2\pi}$.  With 
the equations of motion of this system being:
\begin{eqnarray}
R_{MN}-\frac{1}{2}g_{MN}R=\frac{1}{2}F^2_{MN}-\frac{1}{4}g_{MN}\vert F_4\vert^2
\\
d*F_4+\frac{1}{2}F_4\wedge F_4=0, \qquad dF_4=0. 
\end{eqnarray}
Dimensionally reducing on $S^1$ gives us IIA supergravity. The reduction is specified by:
\begin{eqnarray}
g_{MN}^{11}dx^Mdx^N={\rm e}^{-\frac{2}{3}\Phi}g_{mn}^{10}dx^mdx^n
+{\rm e}^{\frac{4}{3}\Phi}(dx_{10}+C_mdx^m)^2\\
A_{10mn}dx^m\wedge dx^n=\frac{B_{2}}{2\pi R}\qquad A_{lmn}dx^l\wedge dx^m\wedge dx^n=C_{3}\ ,
\end{eqnarray}
where the constant giving the string coupling has been removed from
the dilaton. Then with $2\kappa_0^2g_s^2=\frac{2\kappa_{11}^2}{2\pi
  R}$, where $R$ is the radius of the $X_{10}$ circle on which the
compactification is done, one obtains the IIA supergravity action.

The leading $\alpha'=l_s^2$ low energy effective field theory 
on the dual gravity side is given by IIA supergravity the bosonic part of 
whose action is given in the string frame by:
\begin{equation}
S_{IIA}=\frac{1}{2\kappa_0^2g_s^2} \int d^{10}x\sqrt{-g}\left\{{\rm e}^{-2\Phi}[R+4|d\phi|^2-\frac{1}{12}|H_3|^2-\frac{1}{4}|G_2|^2-\frac{1}{48}|G_4|^2]\right\}+\frac{1}{4\kappa_0^2}\int B_2\wedge dC_3\wedge dC_3\;,\nonumber
\end{equation}
where 
\begin{equation}
H_3=dB_2\;,\qquad G_2=dC_1\;,\qquad G_4=dC_3+H_3\wedge C_1\nonumber \ .
\end{equation}
Eleven dimensional supergravity is the natural strong coupling limit of 
the IIA superstring.
The fields $(\phi,g_{mn},B_{mn})$ are from the $NS\otimes NS$ sector 
of the IIA string while the fields $(C_1,C_3)$ are from the 
$R\otimes R$ sector.

The relevant solution to eleven dimensional supergravity for the dual geometry 
to the BFSS model corresponds to $N$ coincident $D0$ branes in 
the IIA theory. It is given by:
$$
ds^2=-H^{-1}dt^2+dr^2+r^2d\Omega_8^2+H(dx_{10}-C dt)^2$$
with $A_3=0$.

The one-form is given by $C=H^{-1}-1$ and $H=1+\frac{\alpha_0 N}{r^7}$, where 
$\alpha_0=(2\pi)^2 14\pi g_s l_s^7$. After reducing to ten dimensions and taking near horizon limit (with $U=r/\alpha'$) the metric becomes \cite{Itzhaki:1998dd}:
\begin{equation}
ds^2=\alpha'\left(-\frac{F}{\sqrt{H}}dt^2+\frac{\sqrt{H}}{F}dU^2+\sqrt{H}U^2d\Omega_8\right)\ ,
\end{equation}
where $H(U)=\frac{240\pi^5\lambda}{U^7}$ and the black hole time dilation factor is $F(U)=1-\frac{U_0^7}{U^7}$, where $U_0$ is the radius of the horizon related to the Hawking temperature:
\begin{equation}
\frac{T}{\lambda^{1/3}}=\frac{1}{4\pi\lambda^{1/3}}H^{-1/2}F'(U_0)
=\frac{7}{2^415^{1/2}\pi^{7/2}}{\left(\frac{U_0}{\lambda^{1/3}}\right)}^{5/2}.
\end{equation}
Then from black hole thermodynamics the predictions for the entropy and
energy are:
\begin{equation}
\frac{S}{N^2}=\frac{1}{N^2}\frac{A}{4G_N} =  4^{\frac{13}{5}}15^{\frac{2}{5}}\left(\frac{\pi}{7}\right)^{\frac{14}{5}}{\left(\frac{T}{\lambda^{1/3}}\right)}^{9/2}\implies \frac{E}{\lambda^{1/3} N^2}= \left(\frac{2^{21}3^{12}5^2}{7^{19}}\pi^{14}\right)^{1/5}{\left(\frac{T}{\lambda^{1/3}}\right)}^{14/5}
  \end{equation}

\section{Non-perturbative Field Theory Results}
\begin{figure}[H] 
   \centering
   \includegraphics[width=2.9in]{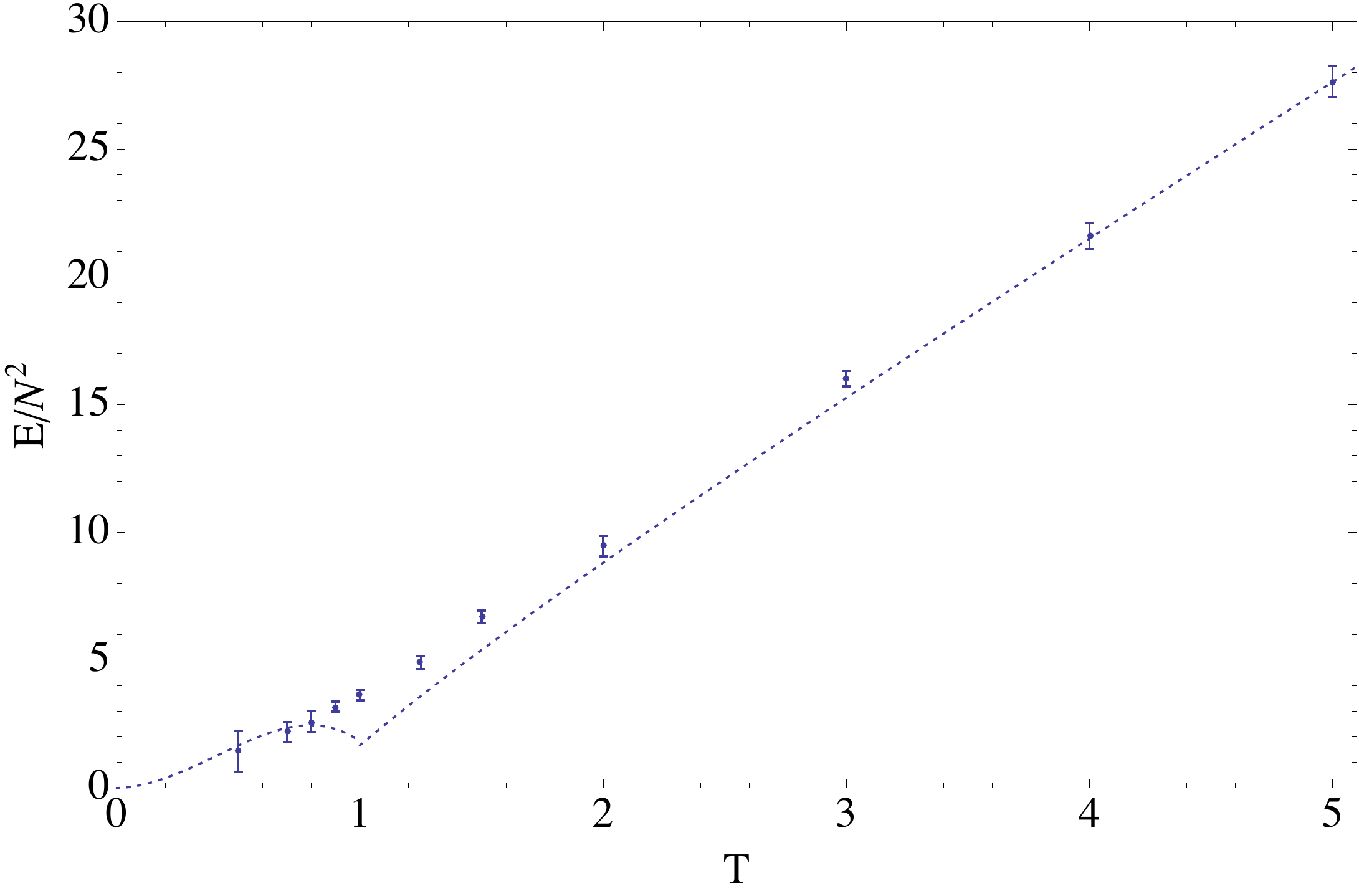} 
   \includegraphics[width=2.9in]{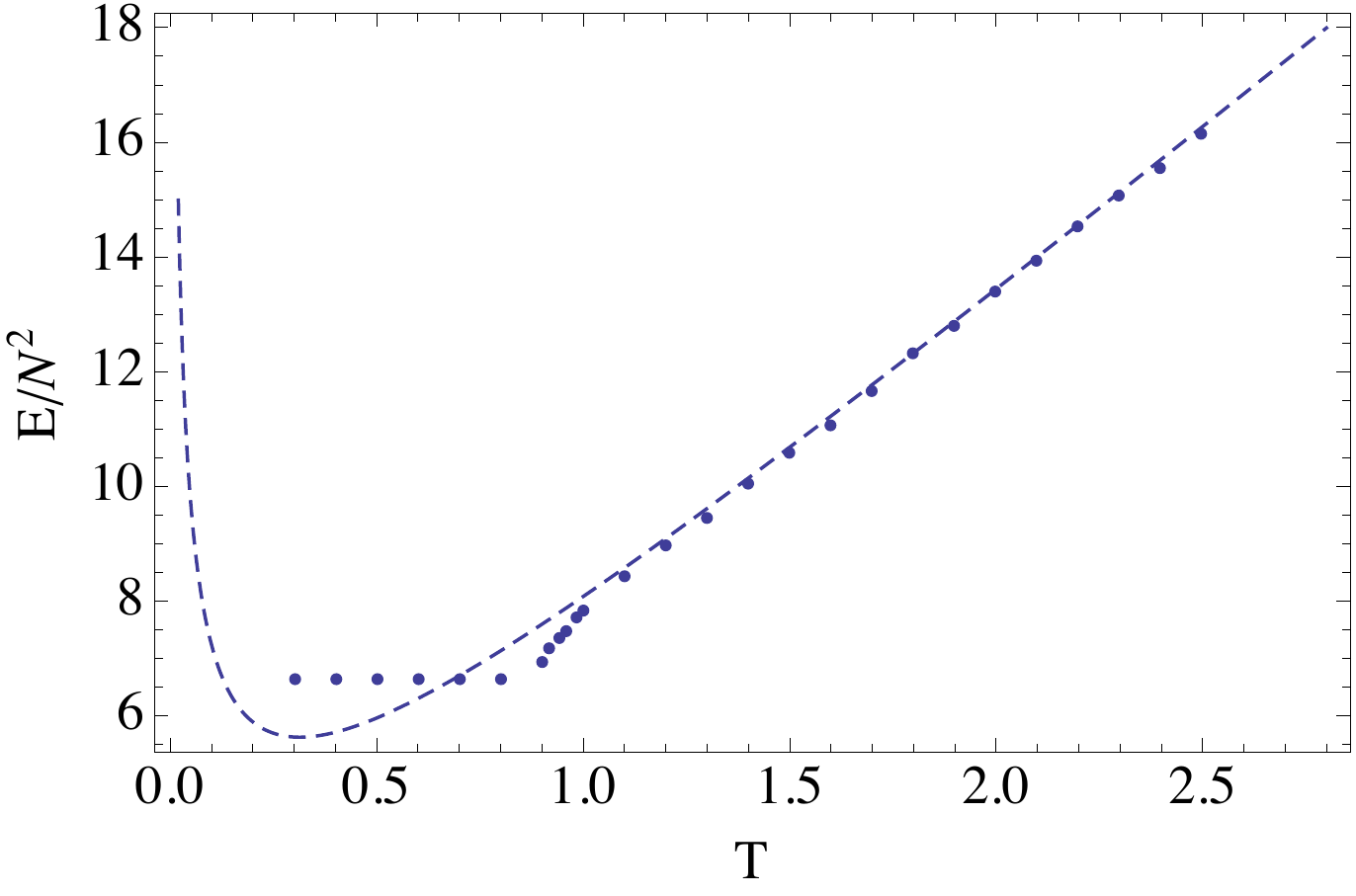} 
   \caption{\small {\it On the left:} Results for the internal energy
   extrapolated from simulations with $8\leq N \leq 14$
   and $8\leq \Lambda \leq 16$. The dashed curve at high temperature
   correspond to the theoretical results of~\cite{Kawahara:2007ib},
   while the low temperature curve represent the prediction for the
   internal energy from the gauge/gravity correspondence~\cite{Hanada:2008ez}. {\it On the right:}
   Plots of the scaled energy $E/N^2$ of the bosonic model
   as functions of the temperature. The dashed curve corresponds to the
   high temperature behaviour obtained in~\cite{Kawahara:2007ib}. One can
   see that near $T\approx 0.9$ the plots suggest the existence of a second
   order phase transition. The energy and temperature in the plots are
   in units of $\lambda^{1/3}$.}
   \label{fig:EnergyComparison}
\end{figure}

To illustrate the difference between the bosonic membrane and its supersymmetric
relative in the eleven dimensional theory we show the expectation value of the
energy, i.e. $\langle H\rangle/N^2$ given by
\begin{eqnarray}
  E/N^2 &=&\left\langle
  -\frac{3}{4 N\beta}
  \int\limits_0^\beta d\tau\,
             {\rm Tr}\left([X^i,X^j]^2\right)\right\rangle
\end{eqnarray}
while for the BFSS model the energy has a fermionic component and is
given by
\begin{eqnarray}
 E/N^2&=&\left\langle
 -\frac{3}{4N\beta} 
 \int_0^\beta d\tau\;
 {\rm Tr} \left([X^i,X^j]^2+\Psi^T C_{10}\,\Gamma^i[X^i,\Psi]\right)\right\rangle  \, .
\end{eqnarray}
\noindent We show the measured values of the two energies in Figure
\ref{fig:EnergyComparison}.

When $1/\alpha'$ corrections are included, the data converges on the low temperature prediction of the 
gauge/gravity correspondence. 
Our results on the BFSS model~\cite{Filev:2015hia} agree well with those of
the other groups that have simulated the system
~\cite{Anagnostopoulos:2007fw}
~\cite{Catterall:2008yz}, 
~\cite{Kadoh:2015mka}.

\section{Conclusions}
\begin{itemize}
\item Bosonic membranes when quantised are massive
  $m\simeq {p}^{1/3} l_p$ and well approximated by a set
  of $p$ massive Gaussian matrix models.
\item Supersymmetric membranes are highly non-trivial with infra-red
  divergences. Gauge/gravity predictions are in excellent agreement
  with non-perturbative tests and the interpretation as a
  non-perturbative formulation of $M$-theory is promising.
\end{itemize}

\acknowledgments
The support from Action MP1405 QSPACE of the COST foundation
is gratefully acknowledged.

\end{document}